\documentclass[10pt, conference, twocolumn]{IEEEtran}

\usepackage{mdwmath}
\usepackage{amsmath}
\usepackage{amssymb}
\usepackage{cite}
\usepackage{epsfig}
\usepackage{url}

\newtheorem{rem}{Remark}
\newtheorem{theo}{Theorem}

\usepackage{graphicx}
\usepackage{graphics}
\usepackage{color}
\usepackage{psfrag}
\usepackage{pstricks, pst-node, pst-tree}
\usepackage{pstricks-add}

\begin{document}
\author{Dimitris S. Papailiopoulos$^{\dagger}$, 
Jianqiang Luo$^{\ddag}$,
Alexandros G. Dimakis$^{\dagger}$,
Cheng Huang$^{*}$, and 
Jin Li$^{*}$\\
$^{\dagger}$University of Southern California, 
Los Angeles, CA 90089, Email:\texttt{\{papailio, dimakis\}@usc.edu}\\
$^{\ddag}$Wayne State University,
Detroit, MI 48202, Email: \texttt{jianqiang@wayne.edu}\\
$^{*}$Microsoft Research, 
Redmond, WA 98052, Email: \texttt{\{cheng.huang, jinl\}@microsoft.com}
}

\title{Simple Regenerating Codes:\\ Network Coding for Cloud Storage}
\maketitle

\begin{abstract}
Network codes designed specifically for distributed storage systems have the potential to provide dramatically higher storage efficiency for the same availability. 
One main challenge in the design of such codes is the exact repair problem: if a node storing encoded information fails, in order to maintain the same level of reliability we need to create encoded information at a new node.
One of the main open problems in this emerging area has been the design of simple coding schemes that allow exact and low cost repair of failed nodes and have high data rates. 
In particular, all prior known explicit constructions have data rates bounded by $1/2$.

In this paper we introduce the first family of distributed storage codes that have simple look-up repair and can achieve arbitrarily high rates.
Our constructions are very simple to implement and perform exact repair by simple XORing of packets. 
We experimentally evaluate the proposed codes in a realistic cloud storage simulator and show significant benefits in both performance and reliability compared to replication and standard Reed-Solomon codes.

\end{abstract}

\section{Introduction}

Distributed storage systems have reached such a massive scale that recovery from failures is now part of regular operation rather than a rare exception~\cite{Sanjay03}.
Large scale deployments typically need to tolerate multiple failures, both for high availability
and to prevent data loss. Erasure coded storage achieves high failure tolerance without requiring
a large number of replicas that increase the storage cost \cite{codvsrep}. Three application contexts where
erasure coding techniques are being currently deployed or under investigation are Cloud storage systems, archival storage, and peer-to-peer storage systems like Cleversafe and Wuala (see e.g.~\cite{KhanBPH,RDP_repair,Survey,storagewiki,Baochun1})

One central problem in erasure coded distributed storage systems is that of maintaining an encoded representation when failures occur.
To maintain the same redundancy when a storage node leaves the system, a {\it newcomer} node has to join the array, access some existing nodes, and exactly reproduce the contents of the departed node. Repairing a node failure in an erasure coded system requires in-network combinations of coded packets, a concept called network coding.
Network coding has been investigated for numerous applications including p2p systems,  wireless ad hoc networks and various storage problems (see e.g. \cite{Gkantsidis,Widmer,Fekri}).

In this paper we focus on network coding techniques for exact repair of a node failure
in an erasure coded storage system~\cite{DimakisGWWR:08}, \cite{storagewiki}.
There are several metrics that can be optimized during repair:
the total information read from existing disks during repair~\cite{ACTEMT,RDP_repair}, the total information communicated in the network \cite{Tamo, PermCodes,RashmiProduct,Salim,Shum,SuhR:09,PDV, KVSKR:09} (called repair bandwidth~\cite{DimakisGWWR:08}), or the total number of disks required for each repair~\cite{Oggier11,KhanBPH}.

Currently, the most well-understood metric is that of repair bandwidth.
For designing $(n,k)$ erasure codes that
have $n$ storage nodes and can tolerate any $n-k$ failures, an information theoretic tradeoff between the repair bandwidth $\gamma$ and the storage per node $\alpha$ was established in~\cite{DimakisGWWR:08}, using cut-set bounds on an information flow graph.
Explicit code constructions exist for the the two extreme points on this bandwidth-storage tradeoff, see e.g.~\cite{Survey,storagewiki}. Despite this substantial amount of prior work, there are no practical code constructions of efficiently repairable
codes with data rates above $1/2$.
Further, different performance metrics might be of interest in different applications.  It seems that for cloud storage applications the main performance bottleneck is the disk I/O overhead for repair, which is proportional to the number of nodes $d$ involved in rebuilding a failed node.

\begin{figure*}[t]
\centerline{\includegraphics[width=1.34\columnwidth]{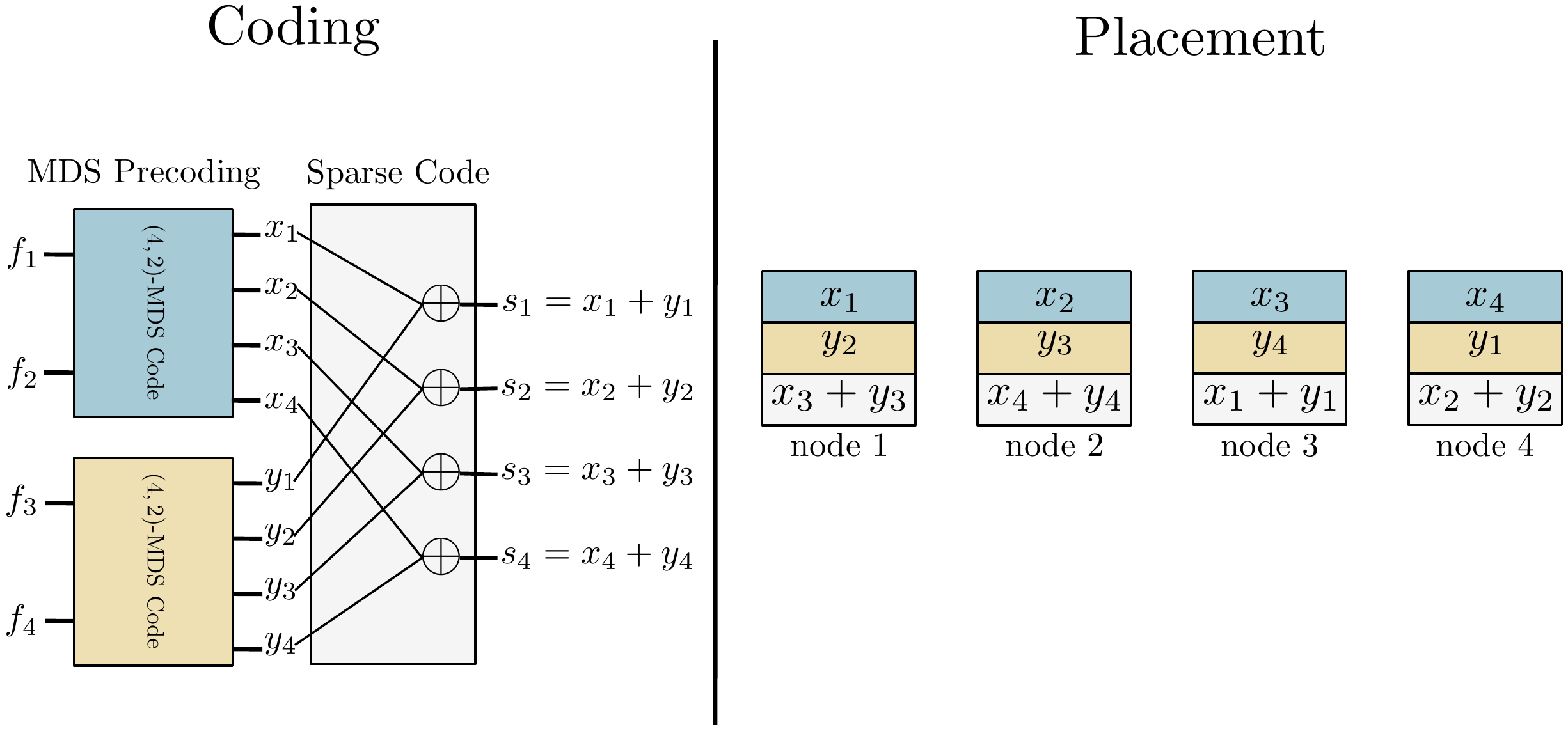}}
\caption{Example of a $(4,2,2)$-SRC. $n=4$ storage nodes, any $k=2$ recover the data and XORs of degree $f=2$ provide simple repair.}
\label{code_42}
\end{figure*}

{\bf Our Contribution}: In this paper we introduce the first family of distributed storage codes that have simple look-up repair and can achieve arbitrarily high rates. Our constructions are very simple to implement and perform exact repair by simple packet combinations. 
Specifically, we design simple regenerating codes (SRC) that have high-rate, very small  disk-I/O $d$, and minimal repair computation. 

An $(n,k,f)$-SRC is a code for $n$ storage nodes that can tolerate $n-k$ erasures, where each node stores a fraction $\frac{f+1}{fk}$ of the file size in coded chunks.
To repair a single coded chunk we need to access $f$ disks and read $1$ chunk from each disk.
The regeneration of an entire lost node costs a fraction $\frac{f+1}{k}$ in repair bandwidth and $d=2f$ disk accesses.
Our codes have rate $R = \frac{f}{f+1}\frac{k}{n}$, which can be made arbitrarily close to $\frac{f}{f+1}$, for constant in $k$ erasure resiliency.

We experimentally evaluate the proposed codes in a realistic cloud storage simulator that models node rebuilds in Hadoop. Our simulator was initially validated on a real Hadoop system of $16$ machines connected by a $1$GB/s network. Our subsequent experiment 
involves $100$ machines and compares the performance of SRC to replication and standard Reed-Solomon codes. We find that SRCs add a new attractive point in the design 
space of redundancy mechanisms for cloud storage.

\section{Simple Regenerating Codes}

The first requirement from our storage code is the $(n,k)$ property: a code will 
be storing information in $n$ storage nodes and should be able \emph{to tolerate any combination of $n-k$ failures} without data loss. 
We refer to codes that have this reliability as ``{\it $(n,k)$ erasure codes},'' or codes that have ``{\it the $(n,k)$ property}.''

One well-known class of erasure codes that have this property is the family of maximum distance separable (MDS) codes~\cite{Survey,evenodd}. In short, an MDS code is a way 
to take a data object of size $M$, split it into chunks of size $M/k$ and create $n$ chunks \emph{of the same size} that have the $(n,k)$ property. 
It can be seen that MDS codes achieve the $(n,k)$ property with the  minimum storage overhead possible: any $k$ storage nodes jointly store $M$ bits of useful information, which is the minimum possible to guarantee recovery. 

Our second requirement is efficient exact repair~\cite{Survey}. When one node fails or 
becomes unavailable, the stored information should be easily reconstructable using 
other surviving nodes. Simple regenerating codes achieve the $(n,k)$ property and simple repair simultaneously by separating the two problems. 
Large MDS codes are used to provide reliability against any $n-k$ failures while very simple XORs applied \emph{over the MDS coded packets} provide efficient exact repair when single node failures happen.

\begin{figure}[h]
\centerline{\includegraphics[width=0.7\columnwidth]{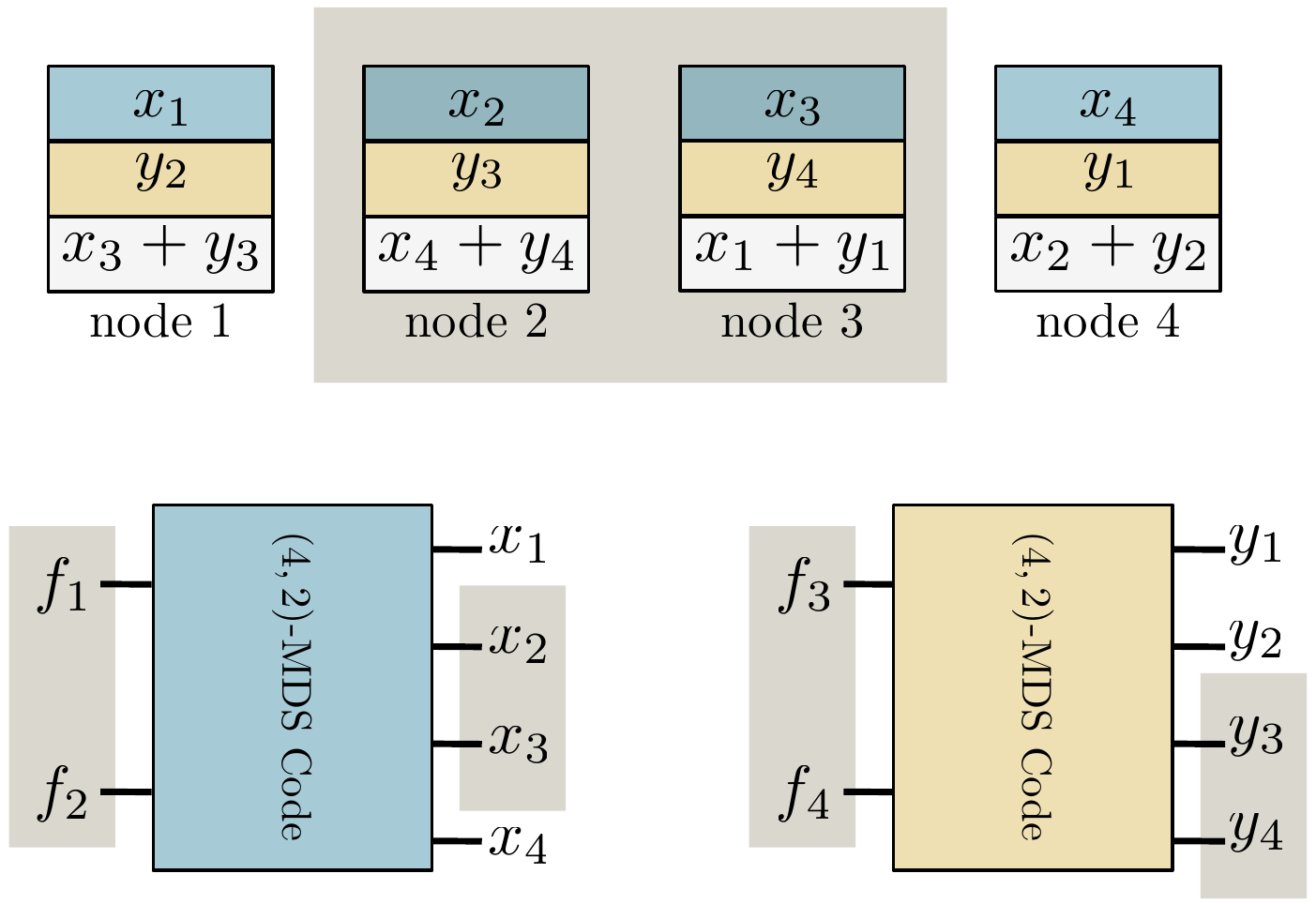}}
\caption{File reconstruction of a $(4,2,2)$-SRC.}
\label{rec_42}
\end{figure}

We give a first overview of our construction through a simple example in Fig. \ref{code_42},
which shows an $(n=4,k=2,f=2)$-SRC.
The original data object is split in $4$ chunks $f_1,f_2,f_3,f_4$.
We first encode $[f_1\;f_2]$ in $[x_1\;x_2\;x_3\;x_4]$ and  
$[f_3\;f_4]$ in $[y_1\;y_2\;y_3\;y_4]$ using any standard $(4,2)$ MDS code. 
This can be easily done by multiplication of the data with the $2\times 4$ generator matrix ${\bf G}$ of the MDS code to form $[x_1\;x_2\;x_3\;x_4]=[f_1\;f_2]{\bf G}$ 
and $[y_1\;y_2\;y_3\;y_4]=[f_3\;f_4]{\bf G}$.
Then we generate a parity out of each ``level'' of coded chunks, i.e., $s_i = x_i+y_i$, which results in an aggregate of $12$ chunks.
We circularly place these chunks in $4$ nodes, each storing $3$, as shown in Fig. 1.

It is easy to check that this code has the $(n,k)$ property and in Fig. \ref{rec_42} we show an example by failing nodes $1$ and $4$. Any two nodes contain two $x_i$ and two $y_i$ chunks which through the outer MDS codes can be used to recover the original data object. We note that the parity chunks are not used in this process, which shows the sub-optimality of our construction. 

\begin{figure}[ht]
\centerline{\includegraphics[width=0.5\columnwidth]{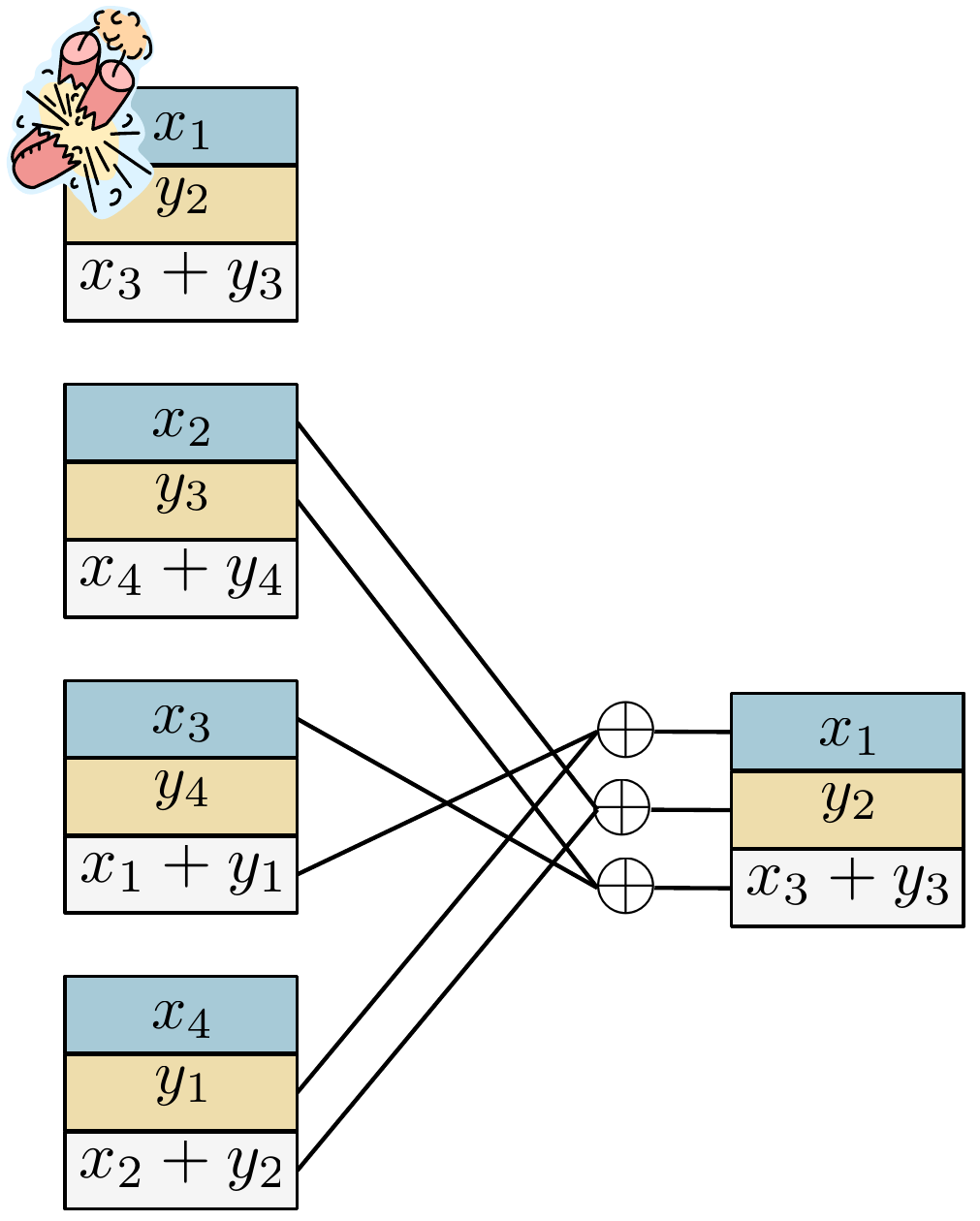}}
\caption{The repair of node $1$ in a $(4,2,2)$-SRC}
\label{rep_42}
\end{figure}

In Fig. \ref{rep_42}, we give an example of a single node repair of the $(4,2,2)$-SRC.
We assume that node $1$ is lost and a newcomer joins the system.
To reconstruct $x_1$, the newcomer has to download $y_1$ and $s_1$ from nodes $3$ and $4$. This simple repair scheme is possible due to the way that we placed the chunks in the $4$ storage nodes: each node stores $3$ chunks with different index.
The newcomer reconstructs each lost chunk by downloading, accessing, and XORing $2$ other chunks. In this process the outer MDS codes are not used.

In short our codes combine outer MDS codes and simple parities to provide fault tolerance and efficient repair respectively. Due to this separation of duties, our codes are suboptimal. However, as we show subsequently this optimality loss corresponds to asymptotically negligible loss in storage efficiency and only a logarithmic factor overhead compared to the optimal information theoretic storage bounds.

\subsection{The $f=2$ Case: degree $2$ parities}

We now present our general SRC construction for the $f=2$ case.
\subsection{Code Construction, Erasure Resiliency, and Rate}
Let a file ${\bf f}$, of size $M = 2k$, that we cut into $2$ parts, say
\begin{equation}
{\bf f} = \left[
{\bf f}^{(1)}\;\;
{\bf f}^{(2)}
\right],
\end{equation}
where ${\bf f}^{(i)}\in\mathbb{F}^{1\times k}$,
for $i\in[2]$, where $[N] = \{1,\ldots,N\}$ and $\mathbb{F}$ is the finite field over which all operations are performed.
Our coding process, is a two-step one: first we independently encode each of the file parts using an outer MDS code and generate simple parity sum out of them.
Then we store the coded chunks and the parity sum chunks in a specific way in $n$ storage components. 
This encode and place scheme enables easy repair of lost chunks and arbitrary erasure tolerance.

We start with an $(n,k)$ MDS code that we use to encode {\it independently} each of the $2$ file parts of size $k$, ${\bf f}^{(1)}$ and ${\bf f}^{(2)}$, into two coded vectors, ${\bf x}$ and ${\bf y}$, of length $n$. 
This encoding process is given by
\begin{equation}
{\bf x} = {\bf f}^{(1)}{\bf G} \text{ and } {\bf y} = {\bf f}^{(2)}{\bf G},
\end{equation}
where  ${\bf G}\in\mathbb{F}^{k\times n}$ is the outer MDS code generator matrix.
We pose no requirements on that MDS code, in the sense that any $(n,k)$ MDS design will work for our purposes.	
The maximum distance of the code ensures that any $k$ encoded chunks of ${\bf x}$  can reconstruct ${\bf f}^{(1)}$; the same goes for any $k$ chunks from ${\bf y}$, i.e., we can use them to reconstruct ${\bf f}^{(1)}$.
We continue by generating a parity sum vector by adding the two coded vectors ${\bf x}$ and ${\bf y}$
\begin{equation}
{\bf s}={\bf x}+{\bf y},
\end{equation}
where $s_l =x_l+y_l$; we note that the index $l$ of the parity sum $s_l$ is the same as the subscript of the $2$ coded chunks that generate it.
This process yields $3n$ chunks: $2n$ coded chunks in the vectors ${\bf x}$ and ${\bf y}$, and $n$ parity sum chunks, i.e., the vector ${\bf s}={\bf x}+{\bf y}$.

We proceed by placing these $3n$ chunks in $n$ storage nodes in the following way: 
each storage node will be storing $3$ chunks, one from ${\bf x}$, one from ${\bf y}$, and one from the parity vector ${\bf s}$.
We require that these $3$ chunks do not share a subscript.
This subscript requirement can be guaranteed by the following circular placement of chunks in the $i$-th node
\begin{equation}
 \left[
\begin{array}{c}
x_i \\
y_{i\oplus1} \\
 s_{i\oplus 2}
\end{array}
\right],
\end{equation}
where $i\in[n]$ and $\oplus$ denotes modulus addition on the ring $\{1,\ldots,n\}$ (for example $n\oplus1 = 1$).
The above circular chunk placement results in the following coded array of $n$ storage nodes
\\
{\small
\begin{align}
\begin{array}{|c|c|c|c|c|c|}
\hline
\text{node }1 & \text{node }2 & \ldots& \text{node } n-2 & \text{node }n-1 & \text{node }n\\
\hline
x_1 & x_2 & \ldots &x_{n-2} & x_{n-1}& x_n  \\
y_2 & y_3 & \ldots &y_{n-1} & y_n & y_1  \\
s_3 & s_4 & \ldots &s_n & s_1 & s_2 \\
\hline
\end{array}\nonumber
\end{align}
}\\
We can observe that for $n\ge2$, indeed the $3$ chunks of each node do not share a subscript.

\subsection{Erasure Resiliency and Effective Coding Rate}

In this section, we present the erasure resiliency and coding rate of the $(n,k,2)$-SRC and prove the following theorem.
Due to lack of space we do not present some proofs in full length and we give sketches instead. 
The extended version of the paper with full proofs can be found online at \cite{extended}.
\vspace{0.21cm}
\begin{theo}
The $(n,k,2)$-SRC can tolerate any possible combination $n-k$ erasures and has effective coding rate $\frac{2}{3}\cdot\frac{k}{n}$.
\end{theo}
\vspace{0.21cm}
{\bf Proof Sketch}:
The $(n,k)$ property of the SRC is inherited by the underlying MDS outer codes:
we can always retrieve the file by connecting to any subset of $k$ nodes of the storage array.
Any subset of $k$ nodes contain $k$ chunks of each of the two file parts
 ${\bf f}^{(1)}$ and ${\bf f}^{(2)}$, which can be retrieved by inverting the corresponding $k\times k$ submatrices of the MDS generator matrix ${\bf G}$
Hence, the $(n,k)$ property of the two identical outer MDS pre-codes renders gives the $(n,k,2)$-SRC its $(n,k)$ property.

We proceed by calculating the coding rate (space efficiency) $R$ of the $(n,k,2)$-SRC, by considering the ratio of the total amount of useful stored information, to the total amount of data that is stored. 
That is, the ratio of the initial file size to the expedited storage
\begin{equation}
R=\frac{\text{file size}}{\text{storage spent}}  = \frac{2\cdot k}{3\cdot n}.
\end{equation}
\hfill$\Box$\\
Hence, the $(n,k,2)$-SRC is an erasure code with rate upper bounded by $\frac{2}{3}$: for fixed erasure tollerance, $n-k=m$, the SRC can have rate arbitrarily close to $\frac{2}{3}$, that is, 
\begin{equation}
\frac{2}{3}\frac{k}{k+m}\overset{k\rightarrow \infty}{\longrightarrow} \frac{2}{3}.
\end{equation}

The $(n,k,2)$ SRC construction that is presented in this section can be generalize to constructions where the rate can be made arbitrarily high.
This is done  by increasing the amount of chunks stored per node and the degree of the parity sums from $2$ to $f$.
These constructions are presented in Section III.

\subsection{Repairing Lost Chunks}

For the general $(n,k,2)$-SRC, when a single node is lost, or a single chunk of that lost node is requested to be accessed, the repair process is initiated.
To sustain high data availability in the presence of chunk and node erasures, the repair process has to be fast and simple: it should be low cost with respect to information read, communicated, and with respect to the number of total disk accesses.
The circular placement of chunks in the SRC enables easy repair of single lost chunks, or single node failures, with respect to the aforementioned metrics.
This is due to the fact that each chunk that is lost shares an index with $2$ more chunks stored in $2$ distinct nodes. 
By contacting these $2$ remaining nodes, we can repair the lost chunk by a simple XOR operation.
For the repair of a single chunk or a single node, we have the following theorem.

\vspace{0.21cm}
\begin{theo}
The repair of a single chunk of the $(n,k,2)$-SRC costs $2$ in repair bandwidth and chunk reads, that is a fraction $\frac{1}{k}$ of the file size, and $2$  disk accesses.
Moreover, the repair of a single node failure costs $6$ in repair bandwidth and chunk reads, that is a fraction $\frac{3}{k}$ of the file size, and $4$ in disk accesses.
\end{theo}
\vspace{0.21cm}
{\bf Proof:}
Let for example node $i\in[n]$ fail, that is, chunks $x_i$, $y_{i\oplus1}$, and  $s_{i\oplus 2}$ are lost.
Then, a newcomer joins the storage array and wishes to regenerate the lost information.
To reconstruct $x^{(1)}_{i}$, the newcomer connects to the two chunks available in the storage system that share the same subscript $i$, i.e., it  connects to the node that contains the parity $s_{i}$ and to the node that contains the chunk $y_{i}$.
The newcomer can then restore the lost chunk $x_i$ simply by subtracting $y_i$ from the parity $s_i$. 
This repair process is summarized in the following $3$ steps.
\begin{center}
{\footnotesize
\begin{tabular}{|c|l|}
\hline
Step &Repair chunk $x^{(1)}_i$:\\
\hline
\texttt{1}&\texttt{Access Disk $i\ominus1$ and download $y_i$}\\
\hline
\texttt{2}&\texttt{Access Disk $i\ominus2$ and download $s_i$}\\
\hline
\texttt{3}&\texttt{restore $x^{(1)}_i:=s_i-x_i$}\\
\hline
\end{tabular}
}
\end{center}
where $\ominus$ is subtraction on the ring $\{1,\ldots,n\}$ (for example $1\ominus1 = n$).
We follow the same manner to repair $y_{i\oplus1}$:
\begin{center}
{\footnotesize
\begin{tabular}{|c|l|}
\hline
Step &Repair chunk $y_{i\oplus1}$:\\
\hline
\texttt{1}&\texttt{Access Disk $i\oplus1$ and download $x_{i\oplus1}$}\\
\hline
\texttt{2}&\texttt{Access Disk $i\ominus1$ and download $s_{i\oplus1}$}\\
\hline
\texttt{3}&\texttt{restore $y_{i\oplus1}:=s_{i\oplus1}-x_{i\oplus1}$}\\
\hline
\end{tabular}
}
\end{center}
The parity repair is also similar, we need to access the $2$ nodes that contain the coded chunks $x_{i\oplus 2}$, and $y_{i\oplus 2}$ and sum them:
\begin{center}
{\footnotesize
\begin{tabular}{|c|l|}
\hline
Step & Repair chunk $s_{i\oplus2}$:\\
\hline
\texttt{1} &\texttt{Access Disk $i\oplus2$ and download $x_{i\oplus2}$}\\
\hline
\texttt{2} &\texttt{Access Disk $i\oplus1$ and download $y_{i\oplus2}$}\\
\hline
\texttt{3} &\texttt{restore $s_{i\oplus2}:=x_{i\oplus2}+y_{i\oplus2}$}\\
\hline
\end{tabular}
}
\end{center}

From the above, we observe that the repair of a single chunk contained in a storage node requires $2$ disk accesses, $2$ chunk reads, and $2$ downloads.
Moreover, to repair a single node failure an aggregate of $6$ chunk reads and $6$ downloads is required.
The set of disks that are accessed to repair all chunks of nodes $i$ is $\{i\ominus2,i\ominus1,i\oplus1,i\oplus2\}$, for $i\in[n]$,
Hence, the number of disk accesses is $\min(n-1,4)$, and $n-1$ is true when $i\ominus2=i\oplus2$, as is the case in our $(4,2,2)$ example in Figures 1-3.
\hfill$\Box$\\

\begin{rem}
We would like to note that a repair would only fail, i.e., one of the packets that are used to regenerate lost information can not be retrieved only if $n\le2$.
\end{rem}

In the following section, we introduce the general code construction of the $(n,k,f)$-SRC, where we consider its rate, reliability, repair properties, and analyze its asymptotics.

\section{SRC: The General Construction}
In this section we generalize the $f=2$ construction, to the $(n,k,f)$-SRC.
For the general $(n,k,f)$-SRC, we use $f$ parallel and identical MDS outer pre-codes and generate a single parity vector from $f$ encoded parts.
We circularly place the generated chunks in $n$ storage nodes.
The $(n,k,f)$-SRC is an $(n,k)$ erasure code with rate $R = \frac{f}{f+1} \frac {k}{n}$, i.e., the SRC always attains a $\frac{f}{f+1}$ fraction of the space efficiency of an $(n,k)$ MDS code, for the same reliability, but with simple and low cost node repair.
We perform single node repairs in the same manner as the $f=2$ case: to repair a chunk, we access $f$ nodes and perform a simple addition.
For any $f$, the communication overhead to repair a single chunk is a fraction $\frac{1}{k}$ of tha file size and the number of chunk reads and disk accesses is $f$, which can be constant and not necessarily a function of $k$.
The repair of a single node failure costs $(f+1)\frac{M}{k}$ in repair bandwidth and we prove that the total number of disk accesses needed for a single node failure is exactly $2\cdot f$.
We proceed by introducing the general code construction and showing its properties.

\subsection{Encoding, Erasure Resilience, and Rate}
Let a file ${\bf f}$, of size $M = fk$, that is subpacketized in $f$ parts,
\begin{equation}
{\bf f} = \left[
{\bf f}^{(1)}
\ldots
{\bf f}^{(f)}
\right],
\end{equation}
 with each ${\bf f}^{(i)}$, $i\in[f]$, having size $k$.
We encode each of the $f$ file parts independently, into vectors ${\bf x}^{(i)}$ of length $n$, using an outer $(n,k)$ MDS code.
That is, we have
\begin{align}
{\bf x}^{(1)} = {\bf f}^{(1)}{\bf G},\;\;\ldots,\;\; {\bf x}^{(f)}& = {\bf f}^{(f)}{\bf G}
\end{align}
where ${\bf G}$ is the $n\times k$ MDS generator matrix.
\begin{rem}
The outer MDS code can be any scalar or array $(n,k)$ MDS code, i.e., we pose no requirements on its design or finite field size.
\end{rem}
We generate a single parity sum vector from all the coded vectors
\begin{equation}
{\bf s} = \sum_{i=1}^{f} {\bf x}^{(i)}.
\end{equation}
This process yields a total of $fn$ coded chunks in the ${\bf x}^{(i)}$ vectors and $n$ parity chunks in ${\bf s}$, i.e., we have an aggregate of $(f+1)n$ chunks available to place in $n$ nodes.

We will circularly place these $(f+1)n$ chunks in $n$ storage nodes, with each node storing $f$ coded chunks and $1$ parity sum chunk, hence each node expends
\begin{equation}
\alpha_{\text{SRC}}=f+1=\frac{f+1}{f} \frac{M}{k}
\end{equation}
in storage capacity.
The placement will again obey the property that enables easy repair: no two chunks within a storage node should share the same subscript.
To ensure successful repair we also require that $f\le n$.
Below we state the circular placement of chunks in the $i$-th node, for $i\in[n]$
\begin{equation}
\left[
\begin{array}{c}
x^{(1)}_i\\
x^{(2)}_{i\oplus1}\\
\vdots\\
x^{(f)}_{i\oplus (f-1)}\\
s_{i\oplus f}
\end{array}
\right],
\end{equation}
which results in the following array of $n$ storage nodes
{\small
\begin{align}
\begin{array}{|c|c|c|c|c|}
\hline
\text{node }1 & \text{node }2 & \ldots& \text{node }n-1& \text{node }n\\
\hline
x^{(1)}_1 & x^{(1)}_2 & \ldots& x^{(1)}_{n-1}& x^{(1)}_n  \\
x^{(2)}_2 & x^{(2)}_3 & \ldots& x^{(2)}_n & x^{(2)}_1  \\
x^{(3)}_3 & x^{(3)}_4 & \ldots& x^{(3)}_1 & x^{(3)}_2  \\
\vdots & \vdots & \ldots  & \vdots& \vdots\\
x^{(f)}_f & x^{(f)}_{f\oplus1} & \ldots & x^{(f)}_{f\oplus(n-2)} & x^{(f)}_{f\oplus(n-1)}  \\
s_{f\oplus1} & s_{f\oplus2} & \ldots & s_{f\oplus (n-1)} & s_{f\oplus n} \\
\hline
\end{array}.\nonumber
\end{align}
}
Then, we have the following theorem.

\vspace{0.21cm}
\begin{theo}
The $(n,k,f)$-SRC can tolerate any combination of $n-k$ node erasures and has coding rate $\frac{f}{f+1}\cdot\frac{k}{n}$.
\end{theo}
\vspace{0.21cm}

{\bf Proof Sketch}:
The $f$ MDS pre-codes guarantee perfect file reconstruction posterior to any $n-k$ erasures.
The file can always be reconstructed by connecting to any $k$ nodes: any collection of $k$ nodes contain $fk$ distinct coded chunks, $k$ of each file part.
Each of these $k$-tuples of coded chunks can give back the information chunks of a single file part due to the $f$ outer MDS codes.

The effective coding rate of the $(n,k,f)$-SRC is equal to the ratio of the initial file size to the expedited storage, that is
\begin{equation}
R_{\text{SRC}} = \frac{\text{file size}}{\text{storage spent}}  = \frac{f\cdot k}{(f+1)\cdot n}.
\end{equation}
\hfill$\Box$\\
By the above theorem we can claim that the rate of the SRC is a fraction $\frac{f}{f+1}$ of the coding rate of an $(n,k)$ MDS code, hence is upper bounded by
\begin{equation}
\frac{f}{f+1}\frac{k}{k+m}\overset{k\rightarrow \infty}{\longrightarrow} \frac{f}{f+1}.
\end{equation}
\hfill$\Box$

\begin{figure}[h]
\centerline{\includegraphics[width=1\columnwidth]{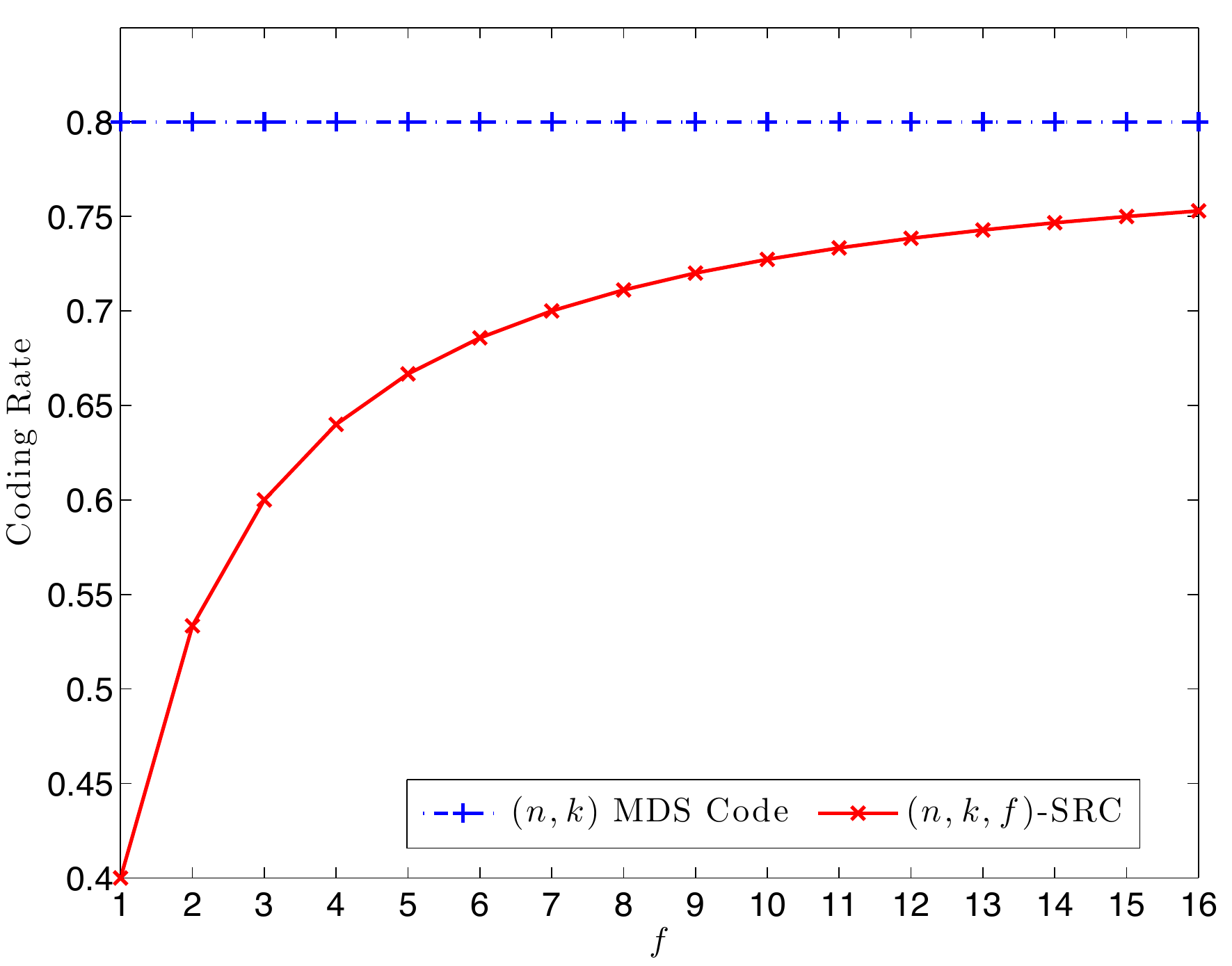}}
\caption{Rate comparison of a $(20,16,f)$-SRC and a $(20,16)$-MDS Code.}
\label{rate_20_16}
\end{figure}

In Fig. \ref{rate_20_16}, we show how the effective coding rate of a $(20,16,f)$ SRC scales as a function of $f$, and compare it with that of a $(20,16)$ MDS code.
Both codes can tolerate $4$ failures. 
We observe that as $f$ increases the coding rate of the SRC approaches that of the MDS code.

\subsection{Repairing Lost Elements}
In this subsection we prove the repair properties of the SRC, which are summarized in the following theorem.

\begin{figure*}[t]
\begin{center}
{\small
\begin{tabular}{||c||c|c|c|c|c||}
\hline
& & & & &\\
& $(n,k)$-MDS & $(n,k,d=n-1)$-MSR & $(n,k,d=k)$-MBR  & $(n,k,d=n-1)$-MBR & $(n,k,f)$-SRC\\
& & & & &\\
\hline
\hline
Storage per node ($\alpha$) & $M/k$ & $M/k$  & $2\frac{k}{k+1}M/k$  & $\frac{2(n-1)}{2(n-1)-k+1}M/k$ & $\frac{f+1}{f}M/k$\\
\hline
Repair Bandwidth ($\gamma$) & $M$ & $\frac{n-1}{n-k}M/k$ & $2\frac{k}{k+1}M/k$  &$\frac{2(n-1)}{2(n-1)-k+1}M/k$ & $(f+1)M/k$\\
\hline
Disk Accesses ($d$) & $k$ & $n-1$ &$k$  &$n-1$ & $2\cdot f$\\
\hline
 Rate ($R$)& $k/n$ & $k/n$&$\frac{1}{2} \cdot\frac{k+1}{n}\le \frac{1}{2}$ & $\le \frac{1}{2}$ & $\frac{f}{f+1}k/n$  \\
\hline
\end{tabular}
}
\end{center}
\caption{$(n,k,f)$-SRC Performance Comparison}
\label{comparison}
\end{figure*}

\vspace{0.21cm}
\begin{theo}
The repair of a single chunk of the $(n,k,f)$-SRC, where each node stores $\alpha_{\text{SRC}} = \frac{f+1}{f}\cdot\frac{M}{k}$, costs $\frac{M}{k}$ in repair bandwidth and $f$ in chunk reads, and disk accesses.
The repair of a single node failure costs
\begin{equation}
\gamma_{\text{SRC}}=(f+1)\frac{M}{k}
\end{equation}
in repair bandwidth, $f(f+1)=(f+1)\frac{M}{k}$ in chunk reads, and
\begin{equation}
d_{\text{SRC}}=\min(2f, n-1)
\end{equation}
in disk accesses.
\end{theo}
\vspace{0.21cm}
{\bf Proof}: Let node $i\in[n]$ fail.
A newcomer node can reconstruct the lost chunk $x^{(l)}_{i\oplus (l-1)}$ by accessing all $f$ nodes in the set
\begin{equation}
\mathcal{S}_{i}(l) = \{i\ominus(f-1+l), i\ominus(f-2+l), \ldots, i\ominus l\}\backslash i.
\end{equation}
and downloading the chunk of each node that has the same subscript $i\oplus(l-1)$ as the lost chunk.
For example to reconstruct $x^{(1)}_{i}$ we need to perform the following steps:
\begin{center}
{\footnotesize
\begin{tabular}{|c|l|}
\hline
Step &Repair chunk $x^{(1)}_{i}$:\\
\hline
\texttt{1}&\texttt{Access Disk $i\ominus 1$ and download $x^{(2)}_{i}$}\\
\hline
\texttt{2}&\texttt{Access Disk $i\ominus 2$ and download $x^{(3)}_{i}$}\\
\hline
\vdots&\vdots\\
\hline
\texttt{f-1}&\texttt{Access Disk $i\ominus (f-1)$ and download $x^{(f)}_{i}$}\\
\hline
\texttt{f}&\texttt{Access Disk $i\ominus f$ and download $s_{i}$}\\
\hline
\texttt{f+1}&\texttt{restore $x_{i}:=s_i-\sum_{l=2}^fx^{(l)}_{i}$}\\
\hline
\end{tabular}
}
\end{center}
Hence, repairing a single coded chunk requires $f=\frac{M}{k}$ chunk downloads, reads, $f$ and disk accesses.
To reconstruct the parity sum chunk $s_{i\oplus f}$, we need to connect to the $f$ nodes that contain the chunks $x^{(l)}_{i\oplus f}$, $l\in[f]$ which generate it.

To repair a single node failure we need to communicate and read $(f+1)f=(f+1)\frac{M}{k}$ symbols.
The total number of disk accesses for a single node repair is given by the number of distinct indices in the set
\begin{equation}
\mathcal{S}_i = \bigcup_{l=1}^{f+1}\mathcal{S}_i(l).
\end{equation}
To enumerate the distinct indices in $\mathcal{S}_i$, we first count the number of distinct indices between sets $\mathcal{S}_i(l)$ and $\mathcal{S}_i(l+1)$ for all $l\in[f]$.
We observe that
{\small
\begin{equation}
\mathcal{S}_i(l)\cup \mathcal{S}_i(l+1) =\{i\ominus(f-1+l), i\ominus(f-2+l), \ldots, i\ominus (l-1)\}\backslash i\nonumber
\end{equation}
}and
\begin{equation}
\left|\mathcal{S}_i(l)\cup \mathcal{S}_i(l+1)\right| =f+1,
\end{equation}
that is, for any two ``consecutive'' chunk repairs, we need to access $f+1$ storage nodes.
Starting with $f$ disk accesses for the first chunk repair, each additional chunk repair requires an additional disk access, with respect to what has already been accessed.
The total number of disks accessed is
{\small
\begin{align}
d_{\text{SRC}}
& = (\text{\# of disks accesses for chunk $x^{(1)}_i$})\nonumber\\
&+  (\text{\# of disks accesses for chunk $x^{(2)}_{i\oplus1}$})\nonumber\\
& \vdots\nonumber\\
&+ (\text{\# of disks accesses for chunk $s_{i\oplus f}$})\nonumber\\
& = f+\underbrace{1+1+\ldots+1}_{f \text{additional disks accesses}}=2\cdot f
\end{align}
}Therefore, to repair a single node failure an aggregate of $2f$ disk accesses is required, when $2f\le n-1$.
If $2f> n-1$ then the number of total disk accesses is $n-1$.
\hfill$\Box$\\

In Fig. \ref{comparison}, we give a comparison table between MDS, MSR, MBR, and Simple Regenerating Codes, with respect to 1) storage capacity per node $\alpha$, 2) repair bandwidth per single node repair $\gamma$, 3) number of disk accesses per single node repair $d$, and 4) effective coding rate $R$.
We consider MSR and MBR codes that connect to $d=\{k,n-1\}$ remaining nodes for a single node failure.
Observe that the number of disk acceses in the SRC is a design parameter that can be set to a constant by appropriately choosing $f$, which can be orders less than $k$. 
\begin{rem}
Regenerating Codes \cite{DimakisGWWR:08} have the property that a single node failure can be repaired by any subset of $d$ remaining nodes, and $k\le d\le n-1$ is fixed by the specific code design.
In sharp contrast, SRCs are look-up repair codes:
for a single node failure, only a specific $d_{\text{SRC}}$ subset of the remaining nodes can reconstruct the file and $d_{\text{SRC}}$ can be a constant, or a function of $k$ that potentially grows much slower than $\Theta(k)$.
\end{rem}

\subsection{Asymptotics of the SRC and links to MDS codes}
In this subsection, we consider the asymptotics of the SRC. What happens if we fix $R=\frac{k}{n}$ and let the degree of parities $f$ grow as a function of $k$?
Let for example
\begin{equation}
f = \log(k).
\end{equation}
Then, the repair of a single node costs $\gamma_{\text{SRC}}=(\log(k)+1)M/k$, with $d_{\text{SRC}}=2f = 2\log(k)$.
In comparison, a single node failure of an $(n,k)$ MSR code costs $\gamma_{\text{MSR}}=\frac{n-1}{n-k} M/k$.
If we let $k$ and $n$ grow and fix $R = \frac{k}{n}$ we obtain
\begin{equation}
\frac{\gamma_{\text{SRC}}}{\gamma_{\text{MSR}}}=\frac{\log(k)+1}{\frac{n-1}{n-k}}=\frac{\log(k)+1}{\frac{1/Rk-1}{(1/R-1)k}}=\Theta(\log(k)).
\end{equation}
The effective coding rate of the SRC is given by
\begin{equation}
\frac{f}{f+1}\frac{k}{n}=\frac{\log(k)}{\log(k)+1}\frac{k}{n}\overset{k \rightarrow \infty }{\longrightarrow}R.
\end{equation}
Therfore, compared to repair optimal MDS codes, i.e. MSR codes, SRCs with $f=\log(k)$ sacrifice asymptotically negligible coding rate and have a logarithmic overhead compared to minimum bandwidth node repair, when at the same time they attain very easy repair based on simple XORs, with logarithmic in $k$ number of disk accesses.

\section{Simulations}

In addition to our theoretical analysis, we evaluate SRCs in a realistic cloud storage simulator. We only tested SRCs with $f=2$ in this paper. This case allows the most efficient repair but at somewhat high storage overhead. We leave the exploration of other choices of $f$ and the involved tradeoffs as future work.

\subsection{Simulator Introduction}
We first present the architecture of the cloud storage system that our simulator is modeling. The architecture contains one master server and a great number of data storage servers, similar to that of GFS~\cite{Sanjay03} and Hadoop~\cite{Konstantin10}. As a cloud storage system may store up to tens of petabytes of data, we expect numerous failures and hence fault tolerance and high availability are critical. To offer high data reliability, the master server needs to monitor the health status of each storage server and detect failures promptly.

In the systems of interest, data is partitioned and stored as a number of fixed-size chunks, which in Hadoop can be 64MB or 128MB. Chunks form the smallest accessible data units and in our system are set to be 64MB. To tolerate storage server failures, replication or erasure codes are employed to generate redundant chunks. Then, several chunks are grouped and form a redundancy set~\cite{Qin03}. If one chunk is lost, it can be reconstructed from other surviving chunks. To repair the chunks due to a failure event, the master server will initiate the repair process and schedule repair jobs.

We implemented a discrete-event simulator of a cloud storage system using a similar architecture and data repair mechanism as Hadoop. To provide accurate simulation results, our simulator models most entities of the involved components such as machines and chunks. When performing repair jobs, the simulator keeps track of the details of each repair process which gives us a detailed performance analysis. 

\subsection{Simulator Validation}
\begin{figure}[ht]
\centering
\includegraphics[width=1\columnwidth]{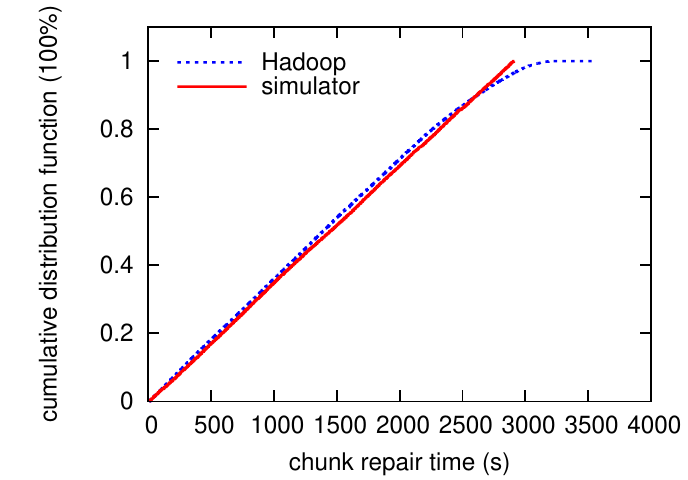}
\caption{CDF of repair time}\label{sim:fig:cdf}
\end{figure}

We first calibrated our simulator to accurately model the data repair behavior of Hadoop. During the validation, we ran one experiment on a real Hadoop system. This system contains 16 machines, which are connected by a 1Gb/s network. Each machine has about 410GB data, namely approximately 6400 chunks. Then, we manually failed one machine, and let Hadoop repair the lost data. After the repair was completed, we analyzed the log file of Hadoop and derived repair time of each chunk. Next, we ran a similar experiment in our simulator. We also collected the repair time of each chunk from the simulation. We present the CDF of the repair time of both experiments in Fig.~\ref{sim:fig:cdf}.

\newcommand{\figurewidth}        {0.45\textwidth}

Fig.~\ref{sim:fig:cdf} shows that the repair result of the simulation matches the results of the real Hadoop system very well, particularly when the percentile is below 95. Therefore, we conclude that the simulator can precisely simulate the data repair process of Hadoop.

\subsection{Storage Cost Analysis}
\begin{figure}[ht]
\centering
\includegraphics[width=1\columnwidth]{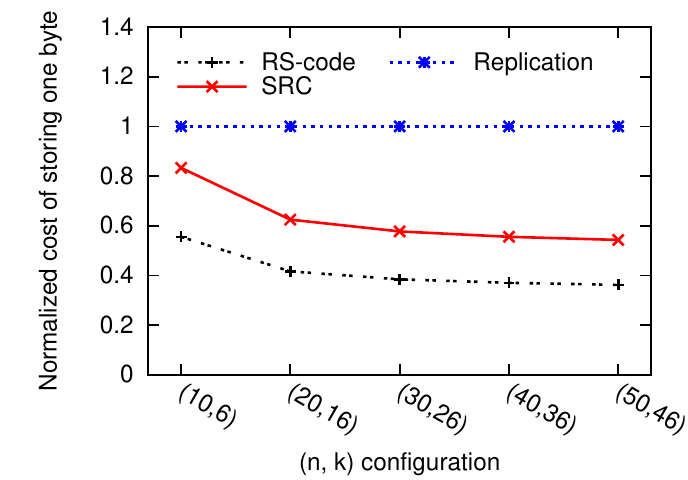}
\caption{Storage cost comparison}\label{sim:fig:cost}
\end{figure}

Now we observe how storage overhead varies when we grow $(n, k)$. We compare three codes: 3-way replication, Reed-Solomon (RS) codes, and SRC. To make the storage overhead easily understood, we define the cost of storing one byte as the metric of how many bytes are stored 
for each useful byte. Obviously, high cost results in high storage overhead. As 3-way replication is a popularly used approach, we use it as the base line for comparison. The result is presented in Fig.~\ref{sim:fig:cost}.

Fig.~\ref{sim:fig:cost} shows that when $n-k$ is fixed, the normalized cost of both the RS-code and the SRC decreases as $n$ grows. When $(n, k)$ grows to $(50, 46)$, the normalized cost of SRC is 0.54, and that of RS-code is 0.36. 
In other words, $(50, 46, 2)$ SRCs need approximately half the storage of 3-way replication. It is worth noting that the cost of SRCs will further reduce if we use larger values of $f$, but at the cost of slower repair. 

\subsection{Repair Performance}~\label{sim:sec:repair}
\begin{figure}[ht]
\centering
\includegraphics[width=1\columnwidth]{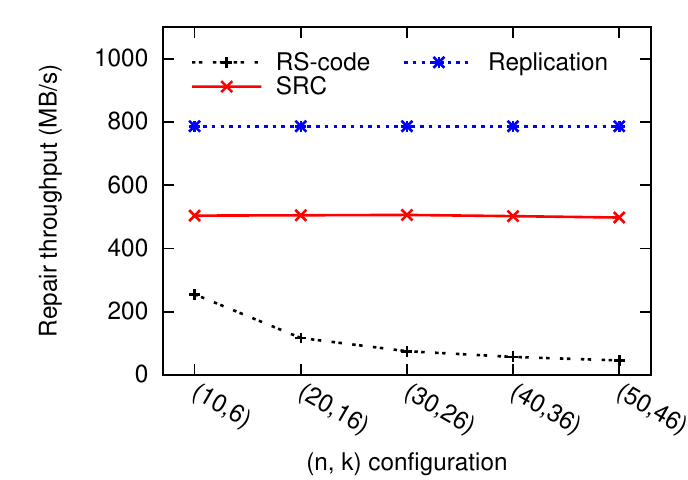}
\caption{Repair performance comparison}\label{sim:fig:repair}
\end{figure}

In this experiment, we measure the throughput of repairing one failed data server. The experiment involves a total of 100 machines, each storing 410GB of data. We fail at random one machine and start the data repair process. After the repair is finished, we measure the elapsed time and calculate the repair throughput. The results are shown in Fig.~\ref{sim:fig:repair}. Note that the throughput of using 3-way replication is constant across different $(n, k)$ since there is no such dependency on these parameters. 

From Fig.~\ref{sim:fig:repair} we can make two observations. First, 3-way replication has the best repair performance followed by SRC, while the RS-code offers the worst performance. This is not surprising due to the amount of data that has to be accessed for the repair. Second, the repair performance of SRC remains constant on various $(n,k)$, but the performance of RS-code becomes much worse as $n$ grows. This is one of the major benefits of SRC, i.e., the repair performance can be independent from $(n, k)$. Furthermore, the repair throughput of SRC is about 500MB/s, approximately 64\% of the 3-way replication's performance.

\subsection{Degraded Read Performance}
In a real system, repair can take place in two situations. One situation is when we need to repair a failed data storage server. Another situation is when we wish to read a piece of data, but it is stored in a storage server that is currently unavailable. The two situations differ in whether the repaired data is stored or not. The first situation is a regular repair operation, which writes the repaired data back to the system. The second situation repairs the data in the main memory and then simply drops it after serving the read request. We call the latter degraded read. The degraded read performance is important, since clients can notice performance degradations when servers have temporary or permanent failures. 

\begin{figure}[ht]
\centering
\includegraphics[width=1\columnwidth]{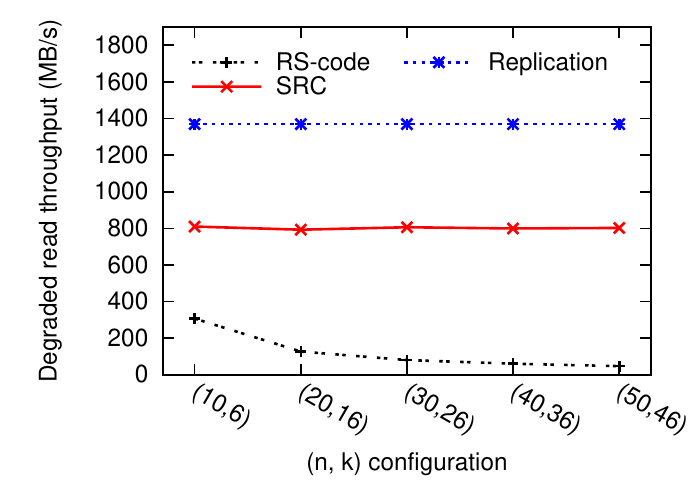}
\caption{Degraded read performance comparison}\label{sim:fig:read}
\end{figure}

We use a similar experimental environment to what we presented in section~\ref{sim:sec:repair}. The only difference is after a chunk is repaired, we do not write it back. The performance results are presented in Fig.~\ref{sim:fig:read}.

We can also make two observations from Fig.~\ref{sim:fig:read}. First, for all three codes, the performance trend of degraded read performance is similar to that of repair performance, shown in Fig.~\ref{sim:fig:repair}. Second, for a code with the same $(n,k)$, the degraded read performance is higher than that of repair performance, due to less accessed data. Again, SRC achieves approximately 60\% degraded read performance of 3-way replication.

\subsection{Data Reliability Analysis}
Now we analyze the data reliability of an SRC cloud storage system. We use a simple  Markov model~\cite{Daniel10} to estimate the reliability. For simplicity, failures happen only to disks and we assume no failure correlations. We note that we expect 
correlated failures to further benefit SRCs over replication since they spread the data to more nodes and hence achieve better diversity protection under correlated failure scenarios. This, however, remains to be verified in a more thorough experimental study of coded cloud storage systems.

\begin{figure}[ht]
\centering
\includegraphics[width=1\columnwidth]{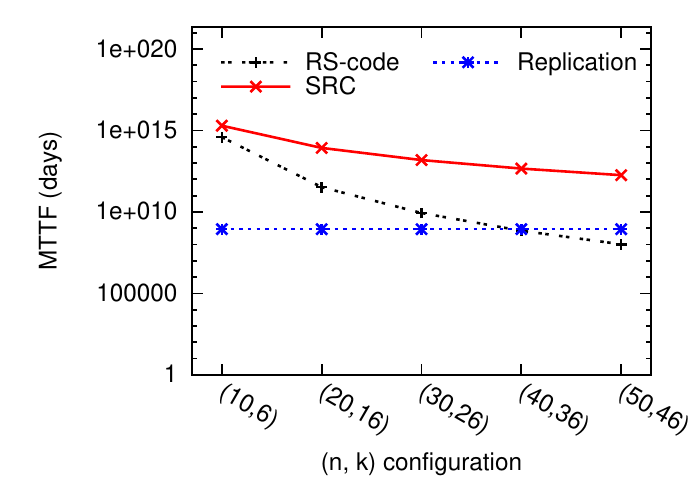}
\caption{MTTF comparison}\label{sim:fig:reliability}
\end{figure}

We assume that the mean time to failure (MTTF) of a disk is 5 years and the system stores 1PB data. To be conservative, the repair time is 15 minutes when using 3-way replication and 30 minutes for SRC, which is in accordance to Fig.~\ref{sim:fig:repair}. In the case of RS-code, the repair time depends on $k$ of $(n, k)$. With these parameters, we first measure the reliability of one redundancy set, and then use it to derive the reliability of the entire system. The estimated MTTF of the entire storage system is presented in Fig.~\ref{sim:fig:reliability}.

Fig.~\ref{sim:fig:reliability} shows that the data reliability of the $3$-way replication is in the order of $10^9$. This is consistent with the results in~\cite{Daniel10}. We can observe that the reliability of SRCs is much higher than 3-way replication. Even for the high rate (low storage overhead) $(50, 46)$ case, SRCs are several orders of magnitude more reliable than 3-way replication. This is benefited from the high repair speed of SRCs. RS codes show a significantly different trend. Although the reliability of $(10, 6)$ and $(20, 16)$ are higher than 3-way replication, the reliability of the RS-code reduces greatly when $(n, k)$ grows. This happens because their repair performance rapidly decreases as $k$ grows.

\section{Conclusions}

We introduced a novel family of distributed storage codes that are formed by combining 
MDS codes and simple locally decodable parities for efficient repair and high fault tolerance. We theoretically show that our codes have the $(n,k)$ reliability, have asymptotically optimal storage and are within a logarithmic factor from optimality in repair bandwidth. One very significant benefit is that the number of nodes that need 
to be contacted for repair can be made a small constant, independent of $n,k$.
Further, SRCs can be easily implemented by combining any prior MDS code implementation with XORing of coded chunks and the appropriate chunk placement into nodes. 

We presented a comparison of the proposed codes with replication and Reed-Solomon codes  using a cloud storage simulator. We have interest on relatively large values of $(n, k)$ because when we keep $n-k$ constant, larger values of $k$ impose lower storage overhead (higher code rates). Standard Reed-Solomon codes cannot operate in this regime since their repair cost increases linearly in $k$. On the contrary, SRCs require only a constant number of nodes involved in each repair and can therefore achieve very good storage overhead with good performance. As an example, if we compare a $(50, 46, 2)$ SRC with 3-way replication we find that the SRC requires approximately half the storage but has approximately 60\% worse degraded read performance. The main strength of the SRC in this comparison,  however, is that it provides approximately four more zeros of data reliability compared to replication. The comparison with Reed-Solomon leads almost certainly to a win of SRCs when slightly more storage is allowed. 

In conclusion we think that SRCs add new feasible points in the tradeoff space of distributed storage codes. They deliver comparable performance to 3-way replication and significantly higher data reliability at a lower storage cost. Our preliminary investigation therefore suggests that SRCs should be attractive for real cloud storage systems.

\end{document}